\documentclass[aps, 
prl, 
twocolumn,
balancelastpage,
amsmath,
amssymb,
floatfix,
longbibliography,
]{revtex4-1}

\usepackage{graphicx}
\usepackage{dcolumn}
\usepackage{bm}

\usepackage{upgreek}     
\usepackage{graphics}     
\usepackage{epsf}       
\usepackage{bm}        
\usepackage{changepage}

\usepackage{hyperref}
\hypersetup{
    colorlinks=true,
    citecolor=magenta,
}

\begin{document}

\title{Cavity probe for real-time detection of atom dynamics in an optical lattice}
\author{R. D. Niederriter}
\author{C. Schlupf}
\author{P. Hamilton}
\email{paul.hamilton@ucla.edu}
\date{\today}
\affiliation{Department of Physics and Astronomy, University of California, Los Angeles, California 90095, USA}

\begin{abstract}
We propose and demonstrate real-time sub-wavelength cavity QED measurements of the spatial distribution of atoms in an optical lattice. 
Atoms initially confined in one ``trap'' standing wave of an optical cavity mode are probed with a second ``probe'' standing wave. With frequencies offset by one free spectral range, the nodes of the trap fall on the anti-nodes of the probe in the ${\approx}$10$^4$ lattice sites around the center of the cavity. 
This lattice site independent atom-cavity coupling 
enables high sensitivity detection of atom dynamics even with atoms spread over many lattice sites.
To demonstrate, we measure the temperature of 20--70~$\mu$K atom ensembles in ${<}$10~$\mu$s by monitoring their expansion by ${\approx}$100~nm after sudden release from the trap lattice. 
Atom-cavity coupling imprints the atom dynamics on the probe transmission. The new technique will enable improved non-destructive detection of Bloch oscillations and other atom dynamics in optical lattices.

\end{abstract}

\maketitle

Force and acceleration sensors based on real-time cavity QED detection of Bloch oscillations of atoms in an optical lattice promise compact size, high bandwidth, and potentially high sensitivity \cite{BlochOscillationsPrasanna2009,BlochOscillationsHolland2009,BlochOscillationsExp2016,BlochOscillationsExp2017}.
These features could enable searches for new short range forces, test certain dark energy models, and be useful in inertial navigation devices.
During Bloch oscillations, the atom wave function periodically stretches and compresses, which leads to variation of the atom-cavity coupling.
In the decade since Bloch-oscillation-based sensors were proposed \cite{BlochOscillationsPrasanna2009,BlochOscillationsHolland2009}, single-shot detection has been observed with short coherence time \cite{BlochOscillationsExp2016,BlochOscillationsExp2017}, but useful sensors have remained out of reach.

Statistical sensitivity of Bloch oscillation based sensors would increase with more atoms, but the atoms must be spread over many lattice sites to prevent collisions and dephasing due to high atom density.
Detecting the atom spatial distribution over many lattice sites requires the atom-cavity coupling to be independent of lattice site. 
The proposed and demonstrated Bloch oscillations experiments \cite{BlochOscillationsPrasanna2009,BlochOscillationsExp2016,BlochOscillationsExp2017} use the trapping lattice to also probe the atom dynamics. While ensuring site-independent coupling, 
this scheme reduces flexibility to independently choose the trap depth and probe detuning; in particular, the trap detuning from the atomic resonance must be small enough to allow sufficient sensitivity for probing but large enough to avoid scattering trap photons. In addition, strong atom-cavity coupling in this case leads to coupling between the atom dynamics and trapping field, which is not always desired.

Separating the trap and probe roles using distinct frequencies enables wide versatility in a cavity QED system.
Typical experiments investigating strong atom-cavity coupling trap the atoms in a far-red-detuned optical lattice and perform measurements with a near-detuned probe frequency \cite{SingleAtomRes_Vuletic2012,StamperKurn,Squeezing_Vuletic2010,SpatiallyHomogeneous_Thompson2016,thompson}.
The difference between trap and probe wavelengths, $\lambda_{t}$ and $\lambda_{p}$ respectively, leads to a limited range over which atoms can be coupled uniformly to the probe. 
The distance (in lattice sites) over which the atom distribution remains nearly uniformly coupled to the probe is $\rm{d}_{\rm{uniform}} \approx \pi/\Delta\phi$, where $\Delta\phi=2\pi (\lambda_{t}-\lambda_{p})/\lambda_{p}$ is the phase shift of the probe standing wave between adjacent sites of the trap standing wave. When using a far-off-resonance trap lattice and a near detuned probe lattice with a typical wavelength difference of ${\approx}$50~nm, $\rm{d}_{\rm{uniform}}<10$. While this can be useful for addressing individual lattice sites \cite{StamperKurn}, it limits the observation of global dynamics such as Bloch oscillations.

\begin{figure}
    \centering
    \includegraphics[width=\columnwidth]{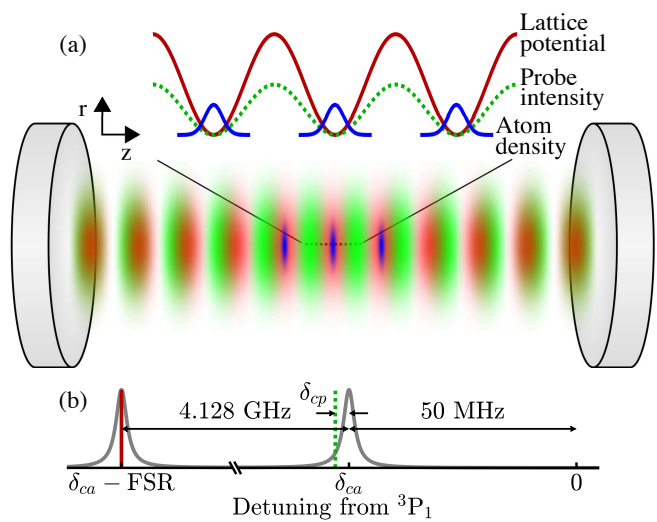}
    \caption{Experimental set-up. 
    (a) Two standing waves of light are formed by driving adjacent longitudinal modes of an optical cavity. The red (solid) line indicates the red-detuned trapping lattice used to confine $^{174}$Yb atoms, shown as blue Gaussian distributions. The green (dotted) line represents a weak probe beam used to measure the spatial extent of the atom density. 
    (b) Detuning of trap lattice (red solid), probe (green dotted), and cavity resonances (black) from the $^3$P$_1$ transition. $\delta_{ca}$ and $\delta_{cp}$ are the detuning of the cavity resonance from the bare atomic transition and the probe beam, respectively.
    } 
    \label{fig:CavityModes}
\end{figure}

In this work, we use a 556~nm trap lattice separated from the probe by one free spectral range (FSR) of $4128.0(1)$~MHz, as shown in Fig.~\ref{fig:CavityModes}. 
Over the ${\approx}$1000 lattice sites (${\approx}$300~$\mu$m) occupied by the atoms, the probe standing wave is shifted by only ${\approx}$1~nm. The coupling uniformity allows observing spatial dynamics of the trapped atoms while remaining far off resonance ($2.3\!\times\! 10^4 \,\Gamma$) compared to the narrow natural linewidth of the Yb intercombination transition, $\Gamma=2\pi\times$180~kHz. 

Site-independent coupling enables extracting global information about the atomic distribution in real time. By time-averaging or probing using adjacent cavity longitudinal modes, others have demonstrated coupling that is independent of atom position \cite{SpatiallyHomogeneous_Thompson2016,Vallet_2017, Hobson:19}. In contrast, in this work, we demonstrate uniform coupling over many lattice sites, but importantly still dependent on the atom distribution within each lattice site. In this way, we can continuously monitor the sub-wavelength motion on microsecond timescales of atoms both bound in and recently released from the lattice.
A related technique sets the probe at a harmonic of the trap lattice, enabling site-independent coupling over the entire cavity, but requires multiple laser wavelengths and more complicated cavity mirror coatings  \cite{Kasevich_SubharmonicTrap_2014,Kasevich_squeezing_SubharmonicTrap_2016,FiberCavity_subHarmonicTrap2018}.

Separating the trap and probe functions for Bloch-oscillation-based force sensors enables fine-tuning the probe without disrupting the trapping lattice. Simulations (following established numerical techniques \cite{BlochOscillationsPrasanna2009}) show orders of magnitude improvement in signal amplitude is possible by independently optimizing the probe detuning. 
We plan to use this scheme in the future with a shallow lattice to observe Bloch oscillations \textit{in situ}. 

Here we take a key step towards high sensitivity detection of Bloch oscillations, observing dynamics of atoms distributed over many lattice sites. 
We derive theoretical expressions for the uniform probe coupling to a trapped atom distribution. We demonstrate an application of this method for a minimally invasive temperature measurement performed in ${<}$10~$\mu$s on ytterbium atoms that begin trapped in an optical lattice and can be recaptured after the measurement.

Consider a cloud of ultracold two-level atoms with atomic resonance frequency $\omega_a$ trapped in a 1-D optical lattice with potential depth $U_t$ and optical frequency resonant with a longitudinal mode of a Fabry-Perot cavity. A probe standing wave is also present with lattice depth $U_p \ll U_t$ and a frequency $\omega_p$, near resonance with another longitudinal mode at $\omega_c$ and nearly one FSR detuned from the trap lattice (see Fig.~\ref{fig:CavityModes}). Interaction with the atoms shifts the cavity resonance and therefore changes the probe transmission through the cavity. 

The atom-cavity system acts as coupled oscillators with normal modes detuned from  atomic resonance by 
\begin{equation}
\Delta\omega_{\pm} = \frac{\delta_{ca}\pm\sqrt{\delta_{ca}^2+\Omega^2}}{2}.
\end{equation} 
The collective vacuum Rabi frequency is $\Omega=2g\sqrt{N \mathcal{I}}$ for $N$ atoms trapped in the lattice with single-atom vacuum Rabi frequency $2g$ and dimensionless atom-probe overlap integral $\mathcal{I}$  \cite{QED}.  The frequency difference between the empty cavity resonance and the atomic resonance is \mbox{$\delta_{ca}=\omega_c-\omega_a$}.
In the far-red-detuned limit where $|\delta_{ca}| \gg \Omega$, 

\begin{equation}
\Delta\omega_-= \frac{\Omega^2}{4\delta_{ca}}.
\end{equation}
In addition to a shift from the atom-cavity coupling, the probe frequency can be detuned, $\delta_{cp}=\omega_c-\omega_p$, from the bare cavity resonance resulting in a total detuning of $\Delta\omega = \Delta\omega_{-} - \delta_{cp}$; see Fig.~\ref{fig:CavityModes}.

The transmitted probe power, $P_{\rm trans}$, at detuning $\Delta\omega$ follows a Lorentzian lineshape with full-width-half-max linewidth $\kappa$,
\begin{equation}
P_{\rm trans}=P_0\frac{1}{1+\left(\frac{\Delta\omega}{\kappa/2}\right)^2},
\label{eq:Lorentzian}
\end{equation}
where $P_0$ is the resonant transmission of the empty cavity. Monitoring the cavity resonance shift thus gives information on the evolution of the atomic spatial distribution as the coupling integral, $\mathcal{I}$, changes (Eq.~\ref{eq:OverlapIntegrals}). 

Next we consider dynamics of the atomic distribution when the trapping lattice is turned off non-adiabatically and the atoms begin to freely expand. Observing the resonance shift of the cavity mode via the probe transmission provides a real-time measurement of the width of the atomic distribution. 
The timescales for changes in transmission due to the axial and radial dynamics are determined by the corresponding spatial scale of the probe beam, which is a TEM$_{00}$ mode of the optical cavity. The atoms are assumed to be confined at the center of the optical cavity over a region much smaller than the Rayleigh length.

The dimensionless coupling integral,
\begin{equation}
\mathcal{I}=\int\rho(\phi,r,z,t)I(\phi,r,z)\mathrm{d}V,
\label{eq:OverlapIntegrals}
\end{equation} 
ranges between 0 and 1 indicating the coupling between the (normalized) atomic spatial distribution $\rho=\rho_{\phi}\rho_r\rho_z$ and the cavity probe spatial intensity profile, $I=I_{\phi}I_rI_z$. The probe intensity and the atomic distribution are cylindrically symmetric and we can factor the coupling integral into radial and axial (z-axis in Fig.~\ref{fig:CavityModes}) contributions, $\mathcal{I}=\mathcal{I}_r\mathcal{I}_z$.
The probe intensity near the waist is
\begin{equation}
    I(r,z) = \sin^2{(k_p z + \phi)}\exp{\left(-2r^2/w^2\right)},
    \label{eq:ProbeIntensity}
\end{equation}
where $w$ is the $1/e^2$ radius of the cavity mode intensity, $k_p=2\pi/\lambda_p$ is the wavenumber of the probe beam, and $\phi$ is the phase shift between the probe and trap lattices. 
The phase $\phi$ enables optimizing the probe overlap for different types of motion. A probe with $\phi=0$ ($\phi=\pm\pi/4$) has maximum sensitivity to changes in the width (center position) of the atom distribution. We focus on measuring the width of the distribution because it is most relevant to detection of Bloch oscillations. 

In the 1-D trapping lattice, the initial positions and velocities of thermalized atoms at each lattice site are well described in the harmonic oscillator limit by Gaussian distributions with standard deviations $\sigma_{i,0}$ in position and $\sigma_{v_i}$ in velocity, where $i=r,z$ refer to the radial and axial directions, respectively. This Gaussian approximation is confirmed for the radial direction by absorption imaging. When the atoms are released from the trapping lattice, the atomic distribution evolves as \cite{Brzozowski}
\begin{equation}
\rho(r,z,t)=\rho_r(r,t)\rho_z(z,t)=\frac{e^{-\frac{r^2}{2\sigma_r^2(t)}}}{\sqrt{2\pi}\sigma_r^2(t)}\frac{e^{-\frac{z^2}{2\sigma_z^2(t)}}}{\sigma_z(t)},
\label{eq:AtomDistribution_rho}
\end{equation}
with $\sigma_i^2(t) = \sigma_{i,0}^2+\sigma_{v_i}^2 t^2$.
We can further specify the velocity standard deviation, $\sigma_{v_i}=\sqrt{k_BT_i/m}$, where $k_B$ is the Boltzmann constant and $T_i$ is the temperature, but choose to leave $\sigma_{i,0}$ as a free parameter.

The atoms start localized at the anti-nodes of the trap lattice and the nodes of the probe lattice (Fig.~\ref{fig:CavityModes}(a)), which we define as $r=z=0$.  Because we have uniform coupling over the ${\approx}$10$^3$ lattice sites that the atoms occupy, the overlap integral for a single lattice site describes the dynamics of the entire ensemble. While below we consider a thermal distribution with no coherence between lattice sites, the full quantum mechanical calculation of the evolution of the spatial distribution is used to simulate Bloch oscillations in the shallow lattice regime, as in \cite{BlochOscillationsPrasanna2009}. 
Thermal expansion provides a simple way to test the new detection scheme since the change in atom distribution between 0 and 1~$\mu$s resembles the stretching of the atom wave function during Bloch oscillations.  
 
Combining Eqs.~\ref{eq:OverlapIntegrals}, \ref{eq:ProbeIntensity}, and \ref{eq:AtomDistribution_rho}, the axial (z-axis) and radial coupling integrals are 
\begin{equation}
\label{eq:Iz}
\mathcal{I}_{z}(t,T_z) =\frac{1}{2}\left(1 - e^{-2k_p^2\sigma_{z}^2(t)}\right) 
\;,
\end{equation}

\begin{equation}
\label{eq:Ir}
\mathcal{I}_r(t,T_r) =\left(1+4\frac{\sigma_r^2(t)}{w^2}\right)^{-1} \;.
\end{equation} 
These overlap integrals determine the cavity resonance shift and produce the time and temperature dependence of the probe beam transmission. 
Because the waist is much larger than the lattice spacing, $w \gg \lambda_p$, the axial overlap $\mathcal{I}_z$ changes much faster than the radial overlap $\mathcal{I}_r$.

\begin{figure}
    \centering
    \includegraphics[width=\columnwidth]{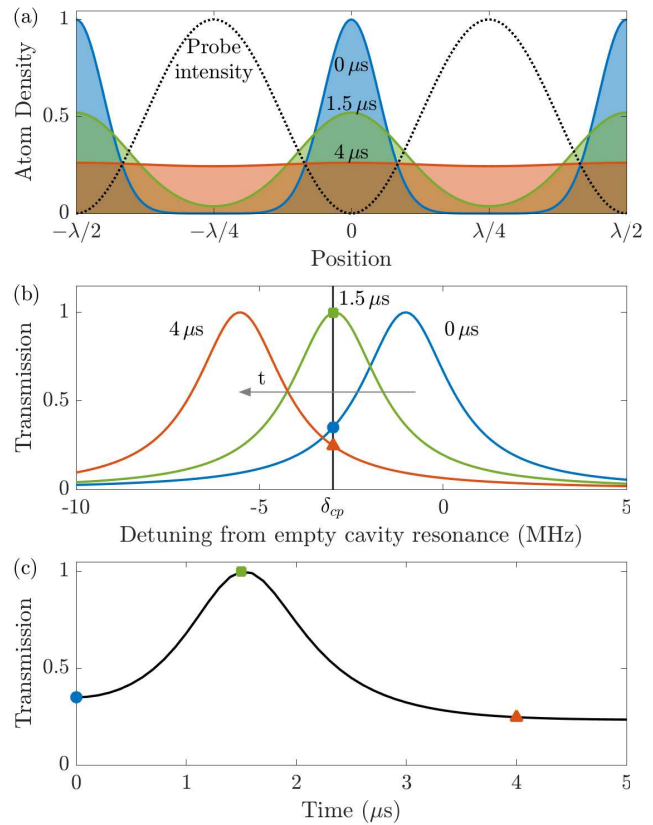}
    \caption{ \label{fig:ProbeOverlapVsTime} Calculated time evolution of the overlap between the atoms at temperature 30~$\mu$K and the probe beam. (a) Atom density at t=0~$\mu$s (blue), 1.5~$\mu$s (green), and 4~$\mu$s (red)  after atoms are released from the trap lattice. The black (dotted) curve shows the 556 nm probe lattice intensity for reference. (b) Cavity transmission spectrum showing the normal mode resonance shift at the same 0, 1.5, and 4~$\mu$s time delays after releasing the atoms from the trap lattice. The vertical line indicates fixed probe detuning at $\delta_{cp}=-2\pi\times3$~MHz, and the three shapes indicate the probe transmission at each time. (c) Probe transmission vs time. Colored shapes correspond to times in part (b).}
\end{figure}

The experiment begins with ${\approx}$10$^6$ $^{174}$Yb atoms cooled to 20--70~$\mu$K and loaded into a 50--200~$\mu$K trap lattice using standard techniques \cite{SupplementalMaterial_All}.
The atom-cavity coupling strength is $2g=2\pi\times48$~kHz. 
A weak probe beam ($U_p<0.05\,U_t$) is coupled into an adjacent TEM$_{00}$ mode of the optical cavity with a linear polarization orthogonal to the trap lattice polarization and red-detuned $\delta_{ca}=-2\pi\times50$~MHz from the atomic resonance. The intensity of the probe beam is kept small to prevent mechanical forces during the free expansion of the atomic cloud.  
The trapping lattice detuning is set one FSR red-detuned from the probe beam, causing the probe and lattice standing waves to be $\pi/2$ out of phase at the center of the cavity (Fig.~\ref{fig:CavityModes}). 

The atoms are released non-adiabatically by switching off the trapping lattice in ${\approx}$100~ns.  They expand for 10~$\mu$s -- 10~ms (Fig.~\ref{fig:ProbeOverlapVsTime}a), 
and the normal mode resonance is shifted by up to several MHz (Fig.~\ref{fig:ProbeOverlapVsTime}b), 
rapidly varying the probe transmission (Fig.~\ref{fig:ProbeOverlapVsTime}c). 
Fast temperature measurements are performed by monitoring the probe transmission through the optical cavity using an  avalanche photodiode (APD) with a bandwidth of 10 MHz. The signal is digitized with a 100 MHz bandwidth oscilloscope and fit with Eq.~\ref{eq:Lorentzian} using Eqs.~\ref{eq:Iz} and \ref{eq:Ir}.

We observe the strong atom-cavity coupling has the predicted effect on the probe beam transmission through the cavity. Qualitatively, the atoms' expansion changes the overlap with the probe mode, which shifts the cavity resonance frequency. The expansion sweeps the cavity resonance frequency across the probe frequency, mapping out the Lorentzian cavity transmission (Eq.~\ref{eq:Lorentzian}). At long time scales, the transmission tends to the empty cavity value as the atoms leave the cavity mode. Quantitatively, we fit the transmission data with the models described above to measure the axial and radial temperatures. The probe beam transmission is recorded for the axial and radial timescales and compared with standard time-of-flight (TOF) temperature measurements. 

\begin{figure}
    \centering
    \includegraphics[width=\columnwidth]{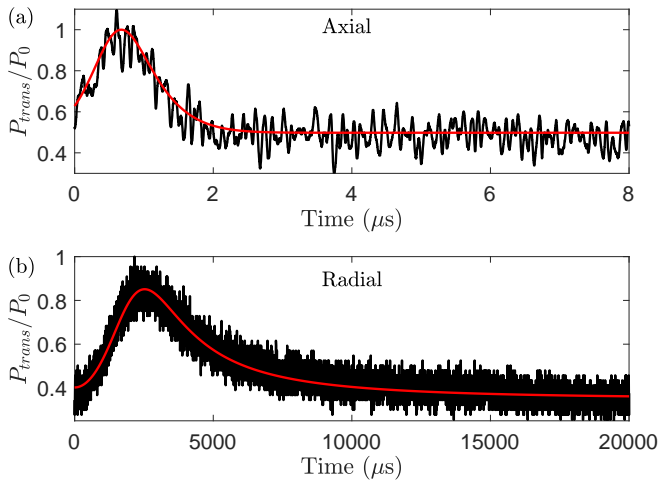}
    \caption{Example normalized raw (un-filtered) traces (black) of probe transmission after releasing the atoms from the trapping lattice. The fits (red) use five free parameters: $P_0, N, \sigma_{i,0}, T_i$, and $\delta_{cp}$. The data were taken with $\delta_{ca}=-2\pi\times50$~MHz, $N=8-10\times10^5$ atoms. The axial trace (a) was taken with $\delta_{cp}=-2\pi\times4$~MHz and TOF $T_z=45~\mu$K, and the radial trace (b) was taken with $\delta_{cp}=-2\pi\times2$~MHz and TOF $T_r=22~\mu$K. Each trace amounts to a single-shot temperature measurement.}
    \label{fig:Traces}
\end{figure}

The transmission function (Eq.~\ref{eq:Lorentzian}) is fit to the measured probe transmission using a non-linear least squares algorithm, with five free parameters: $P_0, N, \sigma_{i,0}, T_i$, and $\delta_{cp}$. The fixed initial probe detuning, $\delta_{cp}$, is included as a free parameter to account for background atoms shifting the resonance by a constant value. Fig.~\ref{fig:Traces} shows example raw data traces of the probe beam transmission observed after release of the atoms. Experimentally using a probe detuning such that the experiment starts and ends on the side of the Lorentzian transmission curve gives the lowest uncertainty.

\begin{figure}
    \centering
    \includegraphics[width=\columnwidth]{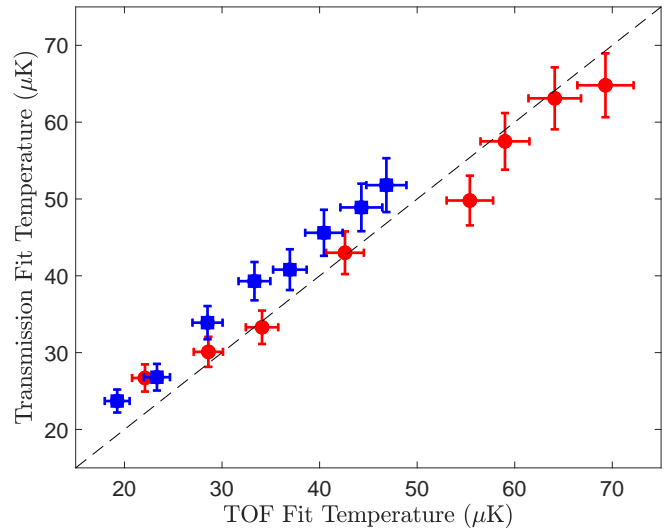}
    \caption{\label{fig:TOFvPEM} Comparison of axial (blue squares) and radial (red circles) temperatures obtained using the new method and the standard time-of-flight technique. The error bars are the quadrature sum of the systematic errors described in the text and the standard error on the mean of ten data points.}
\end{figure}

We vary the initial trap lattice depth and measure the temperature of the atoms using TOF and the new technique. As shown in Fig.~\ref{fig:TOFvPEM}, the new probe measurement obtains similar temperatures to TOF for both axial and radial directions. There are several systematic differences between the two methods. In particular, the probe measurement is sensitive to the coupling of all atoms in the optical cavity mode. Since the signal in the absorption measurement used for TOF is proportional to density, it is less sensitive to a background of diffuse atoms.

The TOF and cavity probe measurements have 5\% and 6\%  systematic error, respectively \cite{SupplementalMaterial_All}.
The statistical error on a TOF temperature measurement comprising seven absorption images is 10\%, leading to an effective single-shot error of 27\%. 
For comparison, the statistical error on a single-shot cavity-probe temperature measurement is 10\% for the axial data and 3\% for the radial data.

We observe a small offset for the axial data between the two methods. We suspect this arises from the fact that the cavity method is more sensitive to the dynamics of a dilute background of atoms that the TOF method misses. The simple assumption that the background distribution of atoms is constant in time is only approximately true; instead, the un-trapped background atoms, which comprise 10--30\% of the total, will retain some periodicity due to the trapping lattice potential. This contributes additional time dependence to the coupling integral, Eq.~\ref{eq:Iz}, when the lattice is turned off. If this population comes from atoms that were heated out of the trap with low radial velocity, they remain overlapped with the probe mode as they fall due to gravity. This population of atoms can be seen in absorption images; however, due to the vertical extent ranging from the trapped atom cloud to the bottom of the optical cavity, we are unable to directly measure their axial temperature with TOF. These atoms will therefore raise the observed temperature of the new measurement without being observed via TOF, consistent with the results in Fig.~\ref{fig:TOFvPEM}. Note that this systematic shift is specific to this application, and will not affect the detection of Bloch oscillations using this technique.

Finally, we note that both methods agree on the total number of atoms as well as the initial cloud width, $\sigma_{r,0}$. Interestingly the initial width inferred from both methods does not agree with a calculation based on a classical phase space distribution for a Gaussian potential. Both the fitting of the probe transmission and the direct measurement of the width via absorption imaging produce a width up to 70\% larger than the calculated width. We expect a broadening of the width when considering the 2-D phase space density including the axial sinusoidal potential, but simulations suggest this only accounts for a 25\% increase. We suspect a more complicated model including a heating term would be necessary to predict the width more accurately, and are interested in continuing to understand this discrepancy and compare observations with other experiments.

We have shown that site-independent coupling between atoms and the cavity mode can be exploited to provide nondestructive real-time information about the spatial distribution of atoms in an optical lattice. We demonstrate one useful application of this by measuring the temperature of atoms confined in an optical lattice. 
The very brief measurement time enables re-capturing the atoms in the lattice. By turning the MOT beams and magnetic field back on after the 10~$\mu$s measurement, about $75\%$ of the atoms are recaptured. Additional cooling should increase this recapture percentage to nearly $100\%$.

Other applications of site-independent coupling include single-shot observation of optomechanical oscillations and heating dynamics; 
based on the results reported here, we infer a sensitivity to center-of-mass oscillations with $\approx$10~nm amplitude. Shifting the probe lattice (setting $\phi=\pi/4$ in Eq.~\ref{eq:ProbeIntensity}) such that the atoms sit on the slope of the probe intensity, the technique would be sensitive to center-of-mass oscillations with amplitude as small as $\approx$1~nm. 
We plan to extend this technique to \textit{in situ} detection of Bloch oscillations \cite{BlochOscillationsPrasanna2009,BlochOscillationsExp2016,BlochOscillationsExp2017}. 
Simulations of Bloch-oscillation-based accelerometers indicate shot-noise-limited sensitivity of $10^{-6}$~g/$\sqrt{\rm Hz}$ with the current experimental parameters, enabling a cm-scale device competitive with portable atom interferometers \cite{AI_review_2019}. 
This sensitivity can be further improved using a narrower cavity linewidth and more atoms. 
This separate trap and probe technique enables optimized detection for Bloch-oscillation-based quantum sensors.


The authors thank Wes Campbell, Eric Hudson, and Holger M{\"u}ller for helpful discussions and acknowledge the generous support of this work by ONR Grant N000141712407 and DARPA Grant D18AP00067. 

R.D. Niederriter and C. Schlupf contributed equally to this work. 

\nocite{ODT_temperature_2000}

\bibliography{CavityProbe} 

\end{document}


\title{Supplemental Material: Cavity probe for real-time detection of atom dynamics in an optical lattice}
\author{R. D. Niederriter}
\author{C. Schlupf}
\author{P. Hamilton}
\email{paul.hamilton@ucla.edu}
\date{\today}
\affiliation{Department of Physics and Astronomy, University of California, Los Angeles, California 90095, USA}

\maketitle

\section{Temperature measurement fit functions}

The cavity probe transmission, $P_{\rm trans}$, is Eq.~3 combined with Eqs.~2 and either 7 (axial) or 8 (radial).
In the axial direction, we use the fit function
\begin{align}
&\frac{P_{\rm trans}}{P_0} = \nonumber\\
&\left(1+\left\{\frac{N g^2 }{\delta_{ca}\kappa}\left[ 1 - e^{-2k_p^2\left(\sigma_{z,0}^2 + \frac{k_B T_z}{m} t^2\right)}\right] \right\}^2\right)^{-1}
\label{eq:AxialFit}
\end{align}
In the radial direction, we use the fit function
\begin{align}
&\frac{P_{\rm trans}}{P_0} = \nonumber\\
&\left(1+\left\{2\frac{N g^2}{\delta_{ca}\kappa} \left[1+\frac{4}{w^2}\left(\sigma_{r,0}^2+\frac{k_BT_r}{m}t^2\right)\right]^{-1} \right\}^2\right)^{-1}
\label{eq:RadialFit}
\end{align}

The fit function includes the single-atom vacuum Rabi frequency $2g = 2\pi\times48$~kHz, the cavity linewidth $\kappa = 2\pi\times3.05(2)$~MHz, the probe wavenumber $k_p=2\pi/\lambda_p$, the $1/e^2$ radius of the TEM$_{00}$ cavity mode $w=140$~$\mu$m, the mass $m$ of $^{174}$Yb, and the Boltzmann constant, $k_B$. 
In the fit, $P_0, N, \sigma_{i,0}, T_i$ and $\delta_{cp}$ are free parameters representing (respectively) the maximum transmitted probe power (on the cavity resonance), the number of atoms, the initial width of the atom distribution, the temperature of the atom distribution, and the detuning of the probe from the cavity resonance. 

\section{Experimental apparatus}

Our apparatus consists of an oven maintained at ${\approx}$450$^\circ$C with five pinhole apertures of 0.5~mm radius (at ${\approx}$550$^\circ$C)
producing an Yb atomic beam that travels down a 35~cm long Zeeman slower operating on the $^1$P$_1$ transition at 399~nm with a linewidth of $2\pi\times29$~MHz; the slower beam detuning is 775~MHz, and affects atoms with velocity up to ${\approx}$350~m/s, which includes the majority of the thermal velocity distribution. The atoms are slowed to ${\approx}$7~m/s and captured in a magneto-optical trap (MOT) using the $^3$P$_1$ transition at 556~nm with a natural linewidth of $2\pi\times180$~kHz (capture velocity ${\approx}$4~m/s). The MOT beams are spectrally broadened to 3~MHz bandwidth using a frequency-modulated acousto-optic modulator (AOM) during the loading phase, then switched to a single frequency to compress the atom cloud. 
The atoms are cooled to ${\approx}$20~$\mu$K, compressed to a cloud of ${\approx}$100~$\mu$m radius, and loaded into the optical trapping lattice formed between a pair of mirrors mounted on a monolithic Zerodur spacer inside the vacuum chamber with spacing controlled by piezo-electric actuators. We typically load $10^6$ $^{174}$Yb atoms in the trapping lattice in 1~s with a single-atom vacuum Rabi frequency of $2g=2\pi\times48$~kHz. 

The detuning of the trapping lattice is controlled via a fiber-coupled waveguide electro-optic modulator (EOM, AdvR) and is several GHz red-detuned from the $^3\mathrm{P}_1$ transition. The cavity free spectral range (FSR) is $2\pi\times4.1280(1)$~GHz with a mirror spacing of 3.63121(9)~cm and a FWHM line width of $\kappa=2\pi \times 3.05(2)$~MHz. 
The cavity length is actively stabilized using the Pound-Drever-Hall (PDH) technique to a modified commercial single-frequency 532~nm laser (Coherent Compass 215M) which is referenced to a hyperfine transition in molecular iodine via modulation transfer spectroscopy.

The trap lattice depth, typically $50-200$~$\mu$K, is calibrated by observation of parametric heating of the atoms when the lattice is modulated at twice its trap frequency. We observe after laser trapping, loading, and thermalization that the axial and radial temperatures of the atoms in the optical lattice is limited by the trap potential depth to $k_B T_z\approx 0.2U_t$ and $k_B T_r\approx 0.3U_t$, similar to the constant ratios observed when loading optical traps \cite{ODT_temperature_2000}.

Images for TOF analysis are obtained in 200~$\mu$s exposures using a 1~cm radius 399~nm absorption beam with saturation parameter $1.7\times10^{-4}$. The beam is directed along the radial axis of the lattice into a CCD camera with a lens resulting in an effective pixel size of 36(1)~$\mu$m. Typically seven absorption images are taken at post-lattice-release times ranging from 0-6~ms, loading new atoms for each image. At each delay time, the radial and axial widths, $\sigma_i(t)$, are determined by Gaussian fits. Fitting the measured widths with $\sigma_i^2(t) = \sigma_{i,0}^2+\sigma_{v_i}^2 t^2$ yields the sample's temperature in each direction.

\section{Measurement uncertainty}

In the TOF measurement, a 3\% uncertainty in the magnification of the imaging system leads to 3\% uncertainty in the temperature. A tilt of the absorption beam with respect to the horizontal axis leads to an effective measured temperature of $T_z^{'}=T_z\cos^2\theta+T_r\sin^2\theta$, where the tilt angle is $\theta=13.1\pm1.8^{\circ}$. The quoted axial temperature corrects for this offset and produces a 1\% uncertainty on $T_z$. We estimate an upper limit to the heating from the absorption beam of ${\approx}$1~$\mu$K at our saturation parameter $1.7\times10^{-4}$. These added in quadrature lead to a 5\% systematic error. The statistical error on a TOF temperature measurement comprising seven absorption images is 10\%, leading to an effective single-shot error of 27\%.

The systematic errors in the new measurement are analyzed by varying aspects of the fit to obtain uncertainties in the temperature. For high probe beam intensity, back-action causes the probe transmission to vary due to atoms oscillating in the probe potential, in agreement with numerical simulations. The intensity of the probe beam is kept low to reduce the frequency and amplitude of these oscillations. Sensitivity to the final background value in numerical testing gives an upper limit of 5\% uncertainty in the temperature from fitting the tail of the data (${\approx}$3-8~$\mu$s in Fig.~3a).
Resolution of the turn-off time of the trapping lattice is limited to $\pm20~$ns uncertainty from a combination of the APD bandwidth and the ring down time of the optical cavity. Changing the start time in the fitting by this amount leads to 3\% change in the temperature. Finally the uncertainty in the background DC voltage of the APD causes a 1\% error on the fit temperature. These combine for a total 6\% systematic error. The statistical error on a temperature measurement is 10\% for the axial data and 3\% for the radial data.

\bibliography{CavityProbe}